\documentstyle[11pt]{article}
\begin{document}
\newcommand{\Si}{\Sigma}
\newcommand{\tr}{{\rm tr}}
\newcommand{\ad}{{\rm ad}}
\newcommand{\Ad}{{\rm Ad}}
\newcommand{\ti}[1]{\tilde{#1}}
\newcommand{\om}{\omega}
\newcommand{\Om}{\Omega}
\newcommand{\de}{\delta}
\newcommand{\al}{\alpha}
\newcommand{\te}{\theta}
\newcommand{\Te}{\Theta}
\newcommand{\vth}{\vartheta}
\newcommand{\be}{\beta}
\newcommand{\la}{\lambda}
\newcommand{\La}{\Lambda}
\newcommand{\D}{\Delta}
\newcommand{\ve}{\varepsilon}
\newcommand{\ep}{\epsilon}
\newcommand{\vf}{\varphi}
\newcommand{\G}{\Gamma}
\newcommand{\ka}{\kappa}
\newcommand{\ip}{\hat{\upsilon}}
\newcommand{\Ip}{\hat{\Upsilon}}
\newcommand{\ga}{\gamma}
\newcommand{\ze}{\zeta}
\newcommand{\si}{\sigma}
\def\bfa{{\bf a}}
\def\bfb{{\bf b}}
\def\bfc{{\bf c}}
\def\bfd{{\bf d}}
\def\bfm{{\bf m}}
\def\bfn{{\bf n}}
\def\bfp{{\bf p}}
\def\bfu{{\bf u}}
\def\bfv{{\bf v}}
\def\bft{{\bf t}}
\def\bfx{{\bf x}}
\def\bfy{{\bf y}}
\def\bfw{{\bf w}}
\newcommand{\li}{\lim_{n\rightarrow \infty}}
\newcommand{\mat}[4]{\left(\begin{array}{cc}{#1}&{#2}\\{#3}&{#4}
\end{array}\right)}
\newcommand{\beq}[1]{\begin{equation}\label{#1}}
\newcommand{\eq}{\end{equation}}
\newcommand{\beqn}[1]{\begin{eqnarray}\label{#1}}
\newcommand{\eqn}{\end{eqnarray}}
\newcommand{\p}{\partial}
\newcommand{\di}{{\rm diag}}
\newcommand{\oh}{\frac{1}{2}}
\newcommand{\su}{{\bf su_2}}
\newcommand{\uo}{{\bf u_1}}
\newcommand{\GL}{{\rm GL}(N,{\bf C})}
\newcommand{\SL}{{\rm SL}(N,{\bf C})}
\def\sln{{\rm sl}(N,{\bf C})}
\newcommand{\gl}{gl(N,{\bf C})}
\newcommand{\PSL}{{\rm PSL}_2({\bf Z})}
\def\f1#1{\frac{1}{#1}}
\newcommand{\rar}{\rightarrow}
\newcommand{\upar}{\uparrow}
\newcommand{\sm}{\setminus}
\newcommand{\ms}{\mapsto}
\newcommand{\bp}{\bar{\partial}}
\newcommand{\bz}{\bar{z}}
\newcommand{\bA}{\bar{A}}
\newcommand{\sect}[1]{\setcounter{equation}{0}\section{#1}}
\renewcommand{\theequation}{\thesection.\arabic{equation}}
\newtheorem{predl}{Proposition}[section]
\newtheorem{defi}{Definition}[section]
\newtheorem{rem}{Remark}[section]
\newtheorem{cor}{Corollary}[section]
\newtheorem{lem}{Lemma}[section]
\newtheorem{theor}{Theorem}[section]

\vspace{0.3in} \vspace{0.3in}
\begin{flushright}
 ITEP-TH-13/01\\
NI01010-SFM\\
\end{flushright}
\vspace{10mm}
\begin{center}
{\Large\bf Solutions of the periodic Toda lattice\\
via the projection procedure
 and by the algebra-geometric method}\\
\vspace{5mm}

M.A.Olshanetsky
\\
{\sf Institute of Theoretical and Experimental Physics, Moscow, Russia,}\\
{\em e-mail olshanet@heron.itep.ru}\\

\vspace{5mm}
\end{center}
\begin{flushright}
{\em Dedicated to 75-th birthday of Vladimir Yakovlevich Fainberg}
\end{flushright}
\vspace{5mm}
\begin{abstract}
In this short review we compare different ways to construct
solutions of the periodic Toda lattice. We give two recipes that follow
from the projection method and compare them with the algebra-geometric
construction of Krichever.
\end{abstract}

\section{Introduction}

\setcounter{equation}{0}
 A large class of classical finite integrable
systems can be derived from free systems by the symplectic reduction
procedure. This class contains the rational and trigonometric  Calogero-Moser
systems \cite{OP1,OP2,KKS}, the open Toda Lattice \cite{OP3,Ko}, elliptic
Calogero-Moser systems
\cite{GN,N}. These systems are particular examples of the Hitchin type
systems. These systems are constructed by the symplectic reduction from a free
two-dimensional field theory defined over basic spectral curves \cite{Hi}.
As by product, this approach allows to construct
the Lax representations and to describe
solutions of the Cauchy problem by the projection method.

 The projection method is working in the following way. Let ${\cal R}$
 will be the upstairs phase space, describing a free
hierarchy and ${\cal R}^{red}$ be the reduced phase space.
It means that there exists a gauge group symmetry ${\cal G}$
of the Hamiltonian hierarchy on
${\cal R}$. Then ${\cal R}^{red}$ is defined in two steps.
First step is  the gauge fixing with
respect to ${\cal G}$. Then one should
 solve the corresponding Gauss law (the moment constraint equation).
The set of its solutions describes ${\cal R}^{red}$.
This procedure is denoted as ${\cal R}^{red}\sim {\cal R}//{\cal G}$.

To solve the Cauchy problem one lifts
the initial data
$x_0\in{\cal R}^{red}$  to a point in the upstairs
space ${\cal R}$ $X_0=\pi^{-1}(x_0)$. The time evolution on ${\cal R}$
 is linear
$$
 X(t_1,t_2,\dots)=X_0+\sum_jF_j(X_0)t_j.
 $$
The solution
$x(t_1,t_2,\dots)$ on ${\cal R}^{red}$ is defined as the gauge
transformation $\pi$ (the projection) of
 $X(t_1,t_2,\dots)$   to a point of the reduced phase space ${\cal R}^{red}$
$$
x(t_1,t_2,\dots)=\pi \left [X(t_1,t_2,\dots)\right ].
$$
The last step is the most difficult part of the projection method, but
for some systems it becomes extremely simple. In particular,
 the rational and trigonometric
Calogero-Moser systems and the open Toda lattice belong this class.
 In these cases the solution of the
Cauchy problem is reduced to the purely algebraic calculations without
quadratures \cite{OP1,OP2,OP3,Ko}. The specific feature of these systems is
the simple dependence of the Lax operator on the spectral parameter, and
as a result, the spectral curves are the rational curves with singularities.

On the other hand there exist algebro-geometric methods based on the finite
gap integration (see, for example, the review \cite{DKN}). For the just
mentioned systems
 the algebra-geometric integration has been proposed fairly
recently\cite{BK,KV}. The procedure
 looks more complicated than the projection
method but leads to the similar expressions.
 For more complicated Lax matrices
the calculation of gauge coinvariant quantities that needed
in the projection method is highly nontrivial problem
and the algebra-geometric methods  become preferable.

Here we compare these two approaches for an intermediate system.
It is the periodic Toda lattice where the dependence on the spectral
parameter is not very complicated. The projection method can be
formulated in terms of
 the Riemann-Hilbert problem or the Gauss decomposition
of  infinite  matrices. On the other hand the Krichever
formula obtained by the algebra-geometric method and
reproduced in \cite{KV,OP} gives an explicit solution to these
problems.

\section{ Periodic Toda lattice as a reduced system}

\setcounter{equation}{0}
 {\bf 1. Definition of the periodic Toda lattice. }\\
The periodic Toda hierarchy describes the one-dimensional $N$-body system
with the coordinates $\bfu=(u_1,\ldots,u_N)$ and the momenta
$\bfv=(v_1,\ldots,v_N)$.  It is a completely integrable
Hamiltonian system with
the canonical symplectic form
 \beq{sf}
 \om=(\de \bfv,\de\bfu)=\sum_{j=1}^N\de v_j\de u_j
\eq
 on the phase space of the Toda lattice
$$
{\cal R}^T=(\bfv,\bfu),~~ \sum_{j=1}^Nv_j=0,~\sum_{j=1}^Nu_j=0.
$$
The second order in momenta Hamiltonian is
\beq{Ha}
H_2=\oh\sum_{j=1}^Nv^2_j+\sum_{j=1}^{N-1}\exp(u_j-u_{j-1})+
\exp(u_N-u_1).
\eq
There are in addition $N-2$ commuting Hamiltonians
providing the complete integrability of the system. They have the form
$$
H_3=\f1{3}\sum_{j=1}^N
v_j^3+v_j(\exp(u_{j+1}-u_j)+\exp(u_{j}-u_{j-1})),~~(u_{N+1}=u_N)
$$
$$
\ldots,
$$
 $$
H_k=\f1{k}\sum_{j=1}^Nv_j^k+\ldots.
$$
If the last term in the potential
(\ref{Ha}) is absent then the corresponding system is called the non-periodic Toda
lattice.

The equations of motion can be represented in the Lax form
\beq{lax}
\p_k=[L,M_k],~~ \p_k=\p_{t_k},
\eq
where $t_k$ is a time corresponding to the evolution with respect to the
Hamiltonian $H_k$ and $L=L(\bfv,\bfu), ~M_k(\bfv,\bfu)$ are matrices of the
$N$-th order depending on the dynamical variables (see below).

\bigskip
\noindent
{\bf 2. Reduction procedure.}\\
Here we describe the reduction procedure for the periodic Toda lattice.
In this way the periodic Toda lattice
Iis described as a result of the symplectic reduction from
a free hierarchy on the cotangent bundle to the loop group $L(\SL)$.
The loop group $L(\SL)$ is the
space  of maps $C^*\to \SL$
$$
L\SL=\{g(z)|,~~g(z)=\sum _\infty^\infty g_az^a,~~g_a\in\SL,~z\in {\bf C}^*\},
 $$
where the only finite number of the matrix coefficients $g_a$ non vanish.
 Let $\Phi$ be a one
form on $C^*$  taking values in Lie$^*(\sln)\sim \sln$. Then the pair
$(\Phi,g)$ parametrizes the cotangent bundle $T^*L(\SL)$. The canonical
symplectic form on $T^*(L\SL)$ is
\beq{2.1}
\om=\f1{2\pi i}\oint_{|z|=1}\tr
(\de(\Phi g^{-1}\de g)\frac{dz}{z}.
 \eq
The form is invariant under the left and right multiplications
\beq{2.2}
\ g\to gf^{(r)}, ~~\Phi\to (f^{(r)})^{-1}\Phi
f^{(r)},~~~f^{(r)}\in L(\SL) ;
\eq
\beq{2.3}
g\to f^{(l)}g,~~\Phi\to \Phi,~~~f^{(l)}\in L(\SL).
\eq

There is a set of gauge invariant Hamiltonians
\beq{ha}
H_k=\f1{k}\oint_{|z|=1}\tr(\Phi^k)\frac{dz}{z}, ~~(k=2,\ldots,N).
\eq
Evidently, these
Hamiltonians are in involution.

The equations of motion for the hamiltonian hierarchy (\ref{2.1}),(\ref{ha})
is
\beq{eq}
\p_j\Phi=0,~~g^{-1}\p_jg=\Phi^{j-1},~~(\p_j=\p_{t_j}),
\eq
 where
 $t_j$ corresponds to the hamiltonian $H_j$. The integration of the
equations gives
 \beq{sol1}
\Phi(t_j)=\Phi_0,~~g(t_2,\ldots,t_N)=
g_0\exp \left(\sum_{j=2}^Nt_j\Phi_0^{j-1}\right).
\eq

Consider in $\SL$ the nilpotent subgroups $N^{(+)} ~(N^{(-)})$ of the
upper (lower) triangular matrices with the units on the diagonal. Let
$L^{(-)}$ be the subgroup of $L(\SL)$
\beq{L-}
L^{(-)}=\{g(z)|_{z\to\infty}\to
n_-\in N^{(-)}\},
\eq
and
\beq{L+}
L^{(+)}=\{g(z)|_{z\to 0}\to n_+\in N^{(+)}\}.
\eq

Since the right action (\ref{2.2}) of $L^{(-)}$ is the canonical
transformations of $\om$ (\ref{2.1}), one can wright down the corresponding
hamiltonian generating (\ref{2.2})
$$
\tr(\Phi\ep), ~~\ep\in {\rm Lie}(L^{(-)}),
$$
 and the moment map
$$
\mu_R~:~T^*L(\SL) \to {\rm Lie}^*(L^{(-)})\sim {\rm Lie}(L^{(+)})
$$
\beq{2.4}
\mu_R(\Phi,g)={\rm Pr}|_{{\rm Lie}(L^{(+)})}(\Phi).
\eq
Similarly,
the left action (\ref{2.3}) of $L^{(+)}$ leads to the
the Hamiltonian
$$
tr(g\Phi g^{-1}\ep), ~~\ep\in {\rm Lie}(L^{(+)}),
$$
and the moment map
\beq{2.5}
\mu_L(\Phi,g)={\rm Pr}|_{{\rm Lie}(L^{(-)})}(g\Phi g^{-1}).
\eq
 We choose the
following values of the momenta
\beq{mu}
 \mu_R=\de_{j,k-1}+z\de_{j,N}\de_{k,1},~~
 \mu_L=\de_{j,k+1}+z^{-1}\de_{j,1}\de_{k,N},
 \eq
 and thereby come to the moment constraints
\beq{2.6}
{\rm Pr}|_{{\rm Lie}(L^{(+)})}(\Phi)=\de_{j,k-1}+z\de_{j,N}\de_{k,1},
\eq
\beq{2.7}
{\rm Pr}|_{{\rm Lie}(L^{(-)})}(g\Phi g^{-1}) =
\de_{j,k+1}+z^{-1}\de_{j,1}\de_{k,N}.
\eq
This choice of the momenta is
minimal, since the whole subgroup $L^{(-)}$ preserves $\mu_R$, while
$L^{(+)}$ preserves $\mu_L$.

For an open dense subset of the loop group $L(\SL)$ one has the Birkhoff
decomposition \cite{PS}.
\beq{bg}
g=r_+hr_-,~~~r_+\in L^{(+)},~r_-\in L^{(-)},~~h
\in~{\rm diagonal~ subgroup}~D\subset \SL.
\eq
Let $U_+,U_-$ be neighborhoods of the points
$z=0$ and $z=\infty$ such that ${\bf C}P^1=U_+\cup  U_-$.
The matrix function $g(z)$ is the transition function of a trivial
holomorphic vector $\SL$ bundle over ${\bf C}P^1$.
The left multiplications of $g(z)$ by the holomorphic
matrix functions on $U_+$ are restricted to those that belong to $L^{(+)}$.
Similarly, the admissible right multiplications are restricted
to the functions from $ L^{(-)}$.
The conditions (\ref{L-}),(\ref{L+})
mean that the bundle has fixed quasi-parabolic structure in the points
$z=0$ and $z=\infty$.
The moduli space of these bundles are described by the elements of the Cartan
subgroup $D$.

As we have mentioned the matrices from $L^{(-)}$ and $L^{(+)}$
 diagonalizing $g$  preserve the moments
 (\ref{mu}). Let us fix the gauge as
\beq{fg}
 h=\di(e^{u_1},\ldots,e^{u_N}),~~\sum_j u_j=0.
 \eq
 Simultaneously, the matrix $\Phi$ is
transformed to the form
\beq{2.8}
\Phi=r_-^{-1}Lr_-.
\eq
Since the moments (\ref{mu}) are invariant under the gauge transform,
one can substitute $h$ and $L$ in the constraints equations (\ref{2.6}),
(\ref{2.7})
$$
{\rm Pr}|_{{\rm Lie}(L^{(+)})}(L)=\de_{j,k-1}+z\de_{j,N}\de_{k,1},
$$
$$
{\rm Pr}|_{{\rm Lie}(L^{(-)})}(hLh^{-1}) =
\de_{j,k+1}+z^{-1}\de_{j,1}\de_{k,N}.
$$
Then $L$ is
defined up to the main diagonal. Its matrix elements can be considered as the moduli of
solutions of the constraint equations
$$
\di (L)=\di(v_1,\ldots,v_N),~~\sum_j v_j=0.
$$
Finally, we come to the following solutions
\beq{2.9}
L_{j,k}=v_j\de_{j,k}+\de_{j,k-1}+\exp(u_j-u_{j-1})\de_{j,k+1}+
\eq
$$
+z\de_{j,N}\de_{k,1}+z^{-1}\exp(u_N-u_1)\de_{j,1}\de_{k,N}.
$$
Therefore, the reduced phase space is described by a pair of matrices
$$ {\cal R}^{red}=T^*(L\SL) //L^{(+)}\oplus
L^{(-)}\sim(L,h).
$$
 It follows from (\ref{fg}) and (\ref{2.9}) that the form
(\ref{2.1}) on ${\cal R}^{red}$ is just the canonical form (\ref{sf}) of the Toda lattice.
 The Hamiltonians
(\ref{ha}) being restricted on $ {\cal R}^{red}$ produce the Hamiltonians
of the Toda hierarchy.
They are in involution, because the original Hamiltonians commute.
Thus the reduced system is the periodic Toda hierarchy.

Substituting (\ref{2.8}) in the first equation (\ref{eq})
we conclude that $L$ satisfies the Lax equation
\beq{l1}
\p_jL=[L,M_j],~~M_j=\p_jr_-r_-^{-1}.
\eq

The matrices $M_j$ can
be  found from the second equation (\ref{eq}). It leads to the
relation between $L$ and $M_j$
$$
h^{-1}r_+^{-1}\p_jr_+h+h^{-1}\p_jh
+M_j=L^{j-1}.
$$
 Since $r_-\in L^{(-)}$, it follows from (\ref{l1}) that $M_j\in{\rm Lie}(L^{(-)})$
 and thereby
\beq{2.10}
 M_j={\rm Pr}|_{{\rm Lie}(L^{(-)})}L^{j-1}.
\eq
 In particular,
$$
M_2=\exp(u_j-u_{j-1})\de_{j,k+1}+z^{-1}\exp(u_1-u_N)\de_{j,1}\de_{k,N}.
 $$
The
matrices $L$ (\ref{2.9}) and $M_j$ (\ref{2.10}) are the Lax pairs for
the periodic Toda hierarchy.

The open Toda lattice can be derived in the similar way if one starts with
the finite-dimensional group $\SL$ instead of the loop group and use
the Gauss decomposition
instead of the Birkhoff decomposition. The corresponding
Lax operator is given by (\ref{2.9}), where the last two terms depending on
the spectral parameter are absent. The $M_j$ operators are expressed through
the $L$ in the same way (\ref{2.10}).

This derivation can be repeat in the same way for any (twisted) affine
algebra $L(G)$.

\section {Solutions via the projection method}
\setcounter{equation}{0}
{\bf 1. Linearization.}\\
 Let $\bfu^0=(u^0_1,\ldots,u^0_N)$ and
$\bfv^0=(v^0_1,\ldots,v^0_N)$ be the Cauchy data for the equations of
motions for the Toda hierarchy. We assume assume that at
$t_j=0,~j=2,\ldots,N$
$$
g(t_2,\ldots,t_N)|_{t_j=0}=h(t_2,\ldots,t_N)|_{t_j=0}=h_0,
$$
$$
\Phi(t_2,\ldots,t_N)|_{t_j=0}=L(t_2,\ldots,t_N)|_{t_j=0}=L_0.
$$
Then,
according to (\ref{sol1}), we have the linear evolution in the upstairs
space
 $$
g(\bft)=X(\bfv^0,\bfu^0,\bft,z),
 $$
 where $\bft=(t_2,\ldots,t_n)$
and
\beq{2.15a}
X(\bfv^0,\bfu^0,\bft,z)=h_0\exp\sum_{j=2}^Nt_jL_0^{j-1}(z).
\eq

The Toda coordinates are obtained from the diagonal part of the Birkhoff
decomposition
(\ref{bg}).
\beq{2.15}
h(\bft)=r_+X(\bfv^0,\bfu^0,\bft,z)r_-
~\rar~\bfu(\bft)= \log(r_+Xr_-).
\eq

The first step in the projection procedure is the calculation
$X(\bfv^0,\bfu^0,\bft,z)$ from the Cauchy data (\ref{2.15a}).
Then one should diagonalize
 $X(\bfv^0,\bfu^0,\bft,z)$ by means of the Birkhoff formula.
 This can be achieved by the two ways
- through the solution of the Riemann problem and by the diagonalization in
the Gauss decomposition of ${\rm GL}(\infty)$. Before describing these methods
we solve the
Cauchy problem for the simpler case - the open Toda lattice.

\bigskip
\noindent
{\bf 2. Solutions of the open Toda Lattice.}\\
 In this case as we
explained the matrix $X(\bfv^0,\bfu^0,\bft)$ does not depend on the spectral
parameter and in (\ref{2.15}) $r_+,r_-$ are the constant triangular matrices
$r_+\in N^{(+)},r_-\in N^{(-)}$. The diagonal elements
$h(\bft)=\di(h_1(\bft),\ldots,h_N(\bft))\in D$ are extracted from the Gauss
decomposition
$$
\SL=N^{(+)}DN^{(-)}.
$$
In fact, the Gauss decomposition is valid for an open dense subset of $\SL$.
Let $\Delta_j(g)$ be the principal
lower minors of order $j,~(j=1,\ldots,N, ~\Delta_N=1)$
of $g\in\SL$. Then this subset is described by the condition
$\Delta_j(g)\neq 0,~j=1,\ldots,N-1$. Moreover, $\Delta_j(g)$
 are invariant with respect to the  multiplication of
$g$ by a matrix from $N^{(+)}$ from the right and by a matrix from $N^{(-)}$
from the left. The relations of the  minors define the diagonal elements
of the Gauss decomposition
$$
h_j=\frac{\Delta_{N-j+1}(X(\bfv^0,\bfu^0,\bft,z))}
{\Delta_{N-j}(X(\bfv^0,\bfu^0,\bft))}.
$$
 Then we obtain
 \beq{ot}
u_j=\log\frac{\Delta_{N-j+1}(X(\bfv^0,\bfu^0,\bft,z))}
{\Delta_{N-j}(X(\bfv^0,\bfu^0,\bft))}.
 \eq
In particular, it follows from (\ref{2.15a}) that the minors
$\Delta_j(\exp X(\bfv^0,\bfu^0,\bft,z))$
are exponential polynomials on times.

\bigskip
\noindent
 {\bf 3. Solutions via the Riemann problem.}\\
Let us rewrite the
Birkhoff decomposition (\ref{bg}) as
\beq{2.16a}
g=b\ti{r}^{-1}_-,
\eq
where
$\ti{r}^{-1}_-$ is a holomorphic matrix outside the contour $|z|=1$ that
satisfies the normalization conditions
\beq{2.16}
\lim_{z\to\infty}\ti{r}_-\to{\rm Id}.
\eq
In other words $\ti{r}_-={\rm Id}+\sum_{j>0}r_jz^{-j}$. The
matrix $b$ is holomorphic inside the contour $|z|=1$. Our ultimate goal is
to find the diagonal, $z$-independent component of $b$. We do it through the
solution of the Riemann problem. Namely, we find first $\ti{r}_-(z)$
starting from $g(z)$. Let us rewrite (\ref{2.16a}) as
$$
b-{\rm Id}=(g-{\rm
Id})\ti{r}_-+\ti{r}_--{\rm Id}.
$$
 Since $b-{\rm Id}$ is holomorphic in $U_+$, $\ti{r}_-$ satisfies the
matrix singular integral equation for $\ti{r}_-$
\beq{ie}
\f1{2\pi i}\oint_{|w|=1}\frac{(g(w)-{\rm
Id})\ti{r}_-(w)}{z-w} +\ti{r}_-(z)-{\rm Id}=0.
\eq
with the normalization condition (\ref{2.16}). Then the solution $\ti{r}_-(z)$
allows to find $b$.

Let
$b=b_0+\sum_{j>0}b_jz^j$. Then $b_0=\lim_{z\to 0}(g\ti{r}_-)$. One can use
the Gauss decomposition of $b_0$ to obtain $h=\di(e^{u_1},\ldots,e^{u_N})$
in (\ref{2.15}), as it was already done for the open Toda lattice
(\ref{ot}).

Substituting in (\ref{ie}) $g(w)=X$ (\ref{2.15a}) we find $\ti{r}_-(z)$,
and then
$$
b(\bfv^0,\bfu^0,\bft,z)=X(\bfv^0,\bfu^0,\bft,z)\ti{r}_-(z)=
h(0,\ldots,0)\exp\left(\sum_{j=2}^Nt_jL_0^{j-1}(z)\right)\ti{r}_-(z).
$$
Finally the solution takes the form
\beq{2.17}
 u_j(\bft)=
 \lim_{z\to 0}\log
 \frac
 {\Delta_{N-j+1}(b(\bfv^0,\bfu^0,\bft,z))}
 {\Delta_{N-j}(b(\bfv^0,\bfu^0,\bft,z))}.
 \eq

\bigskip
\noindent
{\bf 4. Solutions via embedding in ${\rm GL}(\infty)$}.\\
We  define ${\rm GL}(\infty)$ in algebraic terms.
Details can be found in  \cite{PS,UT}.
Let ${\cal H}$ be the Hilbert space of function on $S^1$. The group
${\rm GL}(\infty)$ is subgroup of
all invertible bounded operators in  ${\cal H}$ such that the matrix
elements $M_{j,k}$ of ${\rm GL}(\infty)$ in the Fourier basis $e^{im\vf}$
vanish for large $|j-k|$.

The group $L(\SL)$ can be mapped in  ${\rm GL}(\infty)$.
Consider in ${\rm GL}(\infty)$ the subgroup of periodic matrices
$$
{\rm GL}_{per}(\infty)=\{M_{j+mN,k+mN}=M_{j,k}\}.
$$
Let
$$
g(z)\in L(\SL),~~g(z)=\sum_n(g_{jk})_nz^n,~~(j,k=1\ldots N).
$$
 Then
the image $M\in{\rm GL}_{per}(\infty)$ of $g(z)$ is defined as
$$
M_{s,t}=(g_{jk})_n,~~{\rm for}~s=j+mN,~t=k+(m+n)N,~~s,t,m\in {\bf Z}.
$$

Under this map  the image of $L^{(-)}$ (\ref{L-}) belongs to the
lower triangular matrices in ${\rm GL}_{per}(\infty)$, and
the image of $L^{(+)}$
(\ref{L+}) belongs to the upper triangular matrices. Thus, the Birkhoff
decomposition (\ref{bg})  corresponds to the Gauss decomposition in
${\rm GL}_{per}(\infty)$. The constant diagonal loops give rise
to the periodic diagonal matrices.

Let $\ti{X}(\bfv^0,\bfu^0,\bft)$ be the image of $X(\bfv^0,\bfu^0,\bft,z)$
 (\ref{2.15a}) in ${\rm GL}_{per}(\infty)$, and $\det_j(\ti{X})$ be the
determinant of the semi-infinite matrix $\ti{X}_{m,n},~m,n\in {\bf
Z},~m,n\leq j$. This determinant is bad defined. But relations of them are
well defined.
As it follows
from (\ref{2.15}), the solutions can be expressed through the
logarithm of the fraction
 \beq{2.18}
u_j(\bft)=\log\frac{\Delta_{j+1}(\ti{X})}{\Delta_{j}(\ti{X})}.
 \eq
This
expression is related to the $\tau$-functions of the two-dimensional
Toda hierarchy introduced in \cite{UT} (see below).

Thus we have two implicit formulae for the solutions of the Cauchy problem for
(\ref{2.17}) and (\ref{2.18}) obtained by the projection method.
 We represent in next section the explicit
expression
in terms of the Riemann theta function.

\section{Algebra-geometric calculations}

\setcounter{equation}{0}

Holomorphic vector bundles arise in a natural way in descriptions of
integrable systems. One of this bundles related to the periodic Toda
lattice was described above. We have demonstrated that the
coordinate space is related to the moduli space of the bundle.
The vector bundle can be reconstructed from the spectral characteristics
of the Lax operator.
 Its open a way to
 to construct solutions starting from the spectral data
(see, for example, the review \cite{DKN}).
Here we reproduce the solutions of the periodic Toda lattice constructed by
Krichever \cite{KV,OP}.

 Let ${\cal C}_N$ be the spectral curve defined by
the characteristic polynomial $C_N(\la)$ of the Lax matrix $L$
\beq{SC}
 C_N(\la,z)=0,~~~C_N(\la,z)=\det(L(z)-\la{\rm Id}).
\eq
 It follows from (\ref{2.8}) that (\ref{SC}) is the gauge
invariant equation. Due to the special form of the Lax operator (\ref{2.9})
 the curve is hyperelliptic
 $$
C_N(z,\la)=z+z^{-1}+R_N(\la),~~R_N(\la)=\la^N+\sum_{k=2}^{N}I_k\la^{N-k},
$$
where $I_k$ are another types of the integrals of motion. The curve has genus
$N-1$. It is two-sheets cover of the $\la$-plane with the branching points
$\la_1,\ldots,\la_{2N}$ as the roots of the equation $ R_N(\la)=4$.

The space of holomorphic differentials on ${\cal C_N}$ is generated by
$\la^jd\la/dy,~j=0,\dots,N-2$, and $y=z+\oh R_N(\la)$. Take the linear
combinations of them
$$
\om_i=\sum_{j=0}^{N-2}a_{ij}\la^jd\la/dy,~~i=1,\ldots,N-1 $$ in a such a way
that
$$
2\int_{\la_{2k-1}}^{\la_{2k}}\om_i=\de_{ik},~~k=1,\ldots,N-1,
$$
where the integration is taken over the upper sheet. The differentials
$\om_i$ are the normalized differentials of the first kind. The matrix
$B$ of the $b$-periods
$$
 B_{jk}=2\int_{\la_{2j}}^{\la_{2N}}\om_k
$$
gives rise to the Jacobian ${\cal J}({\cal C_N})$ of the curve ${\cal C_N}$
$$
{\cal J}({\cal C_N})={\bf
C}^{N-1}/(\oplus_j {\bf Z}e_j)\oplus(\oplus_k{\bf Z}B_k),
$$
 $$
e_j=(0,\ldots,\stackrel{j}{1},0\ldots,0),~~B_k=(B_{1k},\ldots,B_{N-1,k}).
 $$
The vector $(\om_1,\ldots,\om_{N-1})$ with the components
$$
\om_k(\ga)=\int_{\la_{2N}}^\ga\om_k
$$
defines the Abel map ${\cal C}_N\to {\cal J}({\cal C}_N)$.

There is  an isomorphism between the submanifold $M_f$ in the phase space
${\cal R}^T$ of levels of integrals and the Jacobian  ${\cal J}({\cal C}_N)$.
 In this
way the Jacobian plays the role of the Liouville torus.
 For a point $(\bfv,\bfu)\in{\cal R}^T$ define $N-1$ points
$\ga_j=(\mu_j,z_j)$ on the spectral curve ${\cal C}_N$. Here $\mu_j$ is a root
of of the characteristic polynomial of the left principal minor of order
$N-1$ of the matrix $L$ (\ref{2.9}), while $z_j$ is a root of (\ref{SC}) at
$\la=\mu_j$, for which the right principle minor of $L$ is non-degenerate.
The isomorphism $M_f\to{\cal J}({\cal C}_N)$ is defined by the vector
$\bfw=(w_1,\ldots,w_{N_1})$
\beq{W}
w_k=-\sum_{n=1}^{N-1}\om_k(\ga_n)+\oh-\oh\sum_{j\neq k}
\int_{\la_{2j-1}}^{\la_{2j}}\om\left(\int_{\la_{2N}}^t\om_k\right)dt.
 \eq

Introduce two differentials
$$
s^1=\frac{\la^{N-1}d\la}{y(z,\la)}+
\sum_{i=0}^{N-2}\al_i\frac{\la^id\la}{y(z,\la)},~~
s^2=\frac{\la^{N}d\la}{y(z,\la)}+
\sum_{i=0}^{N-2}\be_i\frac{\la^id\la}{y(z,\la)}.
$$
They are normalized as
 $$ \int_{\la_{2j-1}}^{\la_{2j}}s^{1,2}=0.
$$
 Let
$S^1,S^2,Z^{\pm}$ be the vectors with coordinates
 $$
S_j^{1,2}=\f1{i\pi}\int_{\la_{2j}}^{\la_{2N}}s^{1,2},~~
Z^{\pm}_j=\int_{\la_{2N}}^{\pm}\om_j.
$$
In the first two integrals the
integration is performed on the upper sheet of the spectral
curve ${\cal C}_N$. In the last integrals the integration is performed on
the upper and lower sheet.

Now we are ready to define the solution of the Cauchy problem for
the evolution
with respect to the quadratic Hamiltonian $H_2$. The initial data
$(\bfv_0,\bfu_0)$ define the spectral curve, the matrix of periods $B$, the
vectors $S^{1,2},Z^\pm$ and ${\bf w}^0={\bf w}(\bfv_0,\bfu_0)$ (\ref{W}).
 Let $\Te(\bfx)$
be the Riemann theta function
 $$
\Te(\bfx)=\sum_{\bfm\in{\bf
Z}^{N-1}}\exp\{i\pi(B\bfm,\bfm) +2i\pi(\bfm\bfx)\},~~B=B_{jk},
$$
Then the
solution is defined as
 \beq{sol}
 u_j(t)=\log\frac {\Te(Z^++\oh
S^2t+S^1(j-1)+\bfw^0)\Te(Z^-+\bfw^0)} {\Te(Z^-+\oh
S^2t+S^1(j-1)+\bfw^0)\Te(Z^++\bfw^0)} +(j-1)C+u_j(0),
\eq
where
$$
Z^\pm=(Z^\pm_1,\ldots,Z^\pm_N),~~S^{1,2}=(S_1^{1,2},\ldots,S_N^{1,2}),
$$
and
the constant $C$ is chosen in a such a way that $u_{N+1}=u_1$.

 This formula can be compared with two expressions obtained above by
 the projection method (\ref{2.17}) and (\ref{2.18}).
  It can be supposed that the relations of the theta functions
  are just the relation of the normalized determinant of the semi infinite matrix
  $\ti {X}$ in (\ref{2.18}).

\section{Other constructions}
\setcounter{equation}{0}

There exist two other approaches describing  solutions of the periodic Toda lattice.
So formally their starting points differ from the projection method they
eventually are formulated in terms of the Riemann problem.

{\bf 1. The tau-functions of the periodic Toda system.}\\
In two papers  \cite{UT,T} the hierarchy of
Toda field theory with infinite numbers of fields was considered.
It was formulated in terms of ${\rm GL}(\infty)$. This hierarchy depends
on two infinite sets of times $\bfx=(x_1,x_2,\dots),~~\bfy=(y_1,y_2,\ldots)$.
 The one-dimensional version depends only on the combinations
 $\ti{t}_2=\oh(x_1-y_1),\ti{t}_3=\oh(x_2-y_2),\ldots$.
TIn this case the infinite Toda hierarchy can be reduced to the finite set of the
Lax equations. In fact, the times $\ti{t}_j$ and related to them operators
$\ti{M}_j$
  differ from (\ref{2.10}). In terms of \cite{UT}
$$
L=B_1+C_1,~~ \ti{M}_{j+1}=B_j-C_j,
$$
where $C_1=M_2=L_-$, $B_1=L_++\di L$.
Consider the following linear problem
\beq{lp}
L V(\bft)=(\mu_L+\mu_R)V(\bft),
\eq
$$
\p_jV(\bft)=\ti{M}_jV(\bft),~~j=2,\ldots,N,~~\p_j=\p_{\ti{t}_j},
$$
where the matrices $\mu_L$ and $\mu_R$ are the moments (\ref{mu}).
The Lax equations for $L$ and $\ti{M}_j$ are the compatibility conditions
for this linear system. The matrix function
$V(\bft)$ can be considered as a matrix wave-function dressing the naked
operators
$\mu_L+\mu_R$ and $\p_j$. Their matrix elements can be expressed as relations
of the tau-functions $\tau(\bfx,\bfy)$, defined in \cite{UT} for the whole
two-dimensional
hierarchy. The tau-functions for the one-dimensional
periodic Toda lattice
is a special subclass of the two-dimensional tau-functions, that
independent on times $\bfx+\bfy$ and satisfy certain periodicity conditions.

The solutions of the Cauchy problem in these terms, proposed in \cite{T}
can be constructed by the following steps.  First, the initial data gives rise
in (\ref{lp})
to the the wave function $V^0=V(\bft=0)$. There exits a representation
for the tau-functions for arbitrary values of times depending on the matrix
elements of $V^0$. The wave-functions $V(\bft)$ are expressed as
relations of $\tau_j(\bft)$. The last step is the solution of the inverse
problem in (\ref{lp}). Thus, the solving of the Cauchy problem can illustrated
by the following scheme
$$
(\bfv^0,\bfu^0)\rar(L^0,\ti{M}_j^0)\to V^0\to \tau(\bft)\to V(\bft)
\to L(\bft).
$$
In these scheme the Riemann problem arises when one constructs the tau functions.

\bigskip
\noindent
{\bf 2. The $R$-matrix method.}\\
As it follows from (\ref{L-}),(\ref{L+}) there exists the decomposition
of the Lie algebra
\beq{3.1}
{\rm Lie}(L(\SL))=
{\rm Lie}(L^{(+)})\oplus {\cal H}\oplus {\rm Lie}(L^{(-)})
={\rm Lie}(B^{(+)})\oplus {\rm Lie}(L^{(-)}),
\eq
where ${\cal H}$ is the diagonal subalgebra of $\SL$ and ${\rm Lie}(B^{(+)})$
is the Borel subalgebra.
Define the projection
$$
M_j^{(+)}=-{\rm Pr}_{{\rm Lie}(B^{(+)})}L^{j-1}.
$$
Then
$$
L^{j-1}=-M_j^{(+)}+M_j^{(-)},
$$
where $M_j^{(-)}=M_j$ (\ref{2.10}).
In other words
$$
M_j^{(\pm)}=-\oh(R\pm 1)L^{j-1},
$$
where
$$
R={\rm Pr}|_{{\rm Lie}(L^{(-)})}-{\rm Pr}_{{\rm Lie}(B^{(+)})}
$$
is the classical $R$-matrix corresponding to the decomposition (\ref{3.1}).
Therefore there exists a pair of the Lax equations for every time $t_k$
\beq{lx}
\p_kL=[L,M_j^{(\pm)}].
\eq
In fact, (\ref{3.1}) corresponds to the Birkhoff decomposition (\ref{2.16a}).
The solutions of the Lax hierarchy (\ref{lx}) can be expressed in terms
of the Birkhoff decomposition
and thereby we come again to the Riemann problem discussed in Section 2.

For
general Lax equations defined on ${\bf C}P^1$ the interrelations
between the classical $R$ matrices, the Riemann problem and the
algebra-geometric methods were discussed in \cite{RS}.

{\bf Acknowledgments.}\\
{\sl I am grateful to the Isaac Newton Institute for Mathematical Studies
(Cambridge) for hospitality where this work was completed. I would like to
thank V.Novokshenov for useful discussions.
The work  is supported in part by grants RFFI-00-02-16530,
 INTAS 99-01782  and 96-15-96455 for support of scientific schools.
}

\small{

}
\end{document}